\documentclass[showpacs,preprintnumbers,twocolumn,amsmath,amssymb]{revtex4}%
\usepackage{epsfig}
\usepackage{subfigure}
\usepackage{graphicx}
\usepackage{dcolumn}
\usepackage{stmaryrd}
\usepackage{mathrsfs}
\usepackage{pifont}
\usepackage{amsthm}
\usepackage{bm}
\usepackage{latexsym}
\usepackage{amsmath}
\usepackage{amsfonts}
\usepackage{amssymb}
\usepackage{mathrsfs,amssymb}%
\providecommand{\U}[1]{\protect\rule{.1in}{.1in}}
\providecommand{\U}[1]{\protect\rule{.1in}{.1in}}

\newcommand{\beq}{\begin{equation}}
\newcommand{\eeq}{\end{equation}}
\newcommand{\beqa}{\begin{eqnarray}}
\newcommand{\eeqa}{\end{eqnarray}}

\begin{document}
\title{Fermionic Stochastic Schr\"{o}dinger Equation and Master Equation: An Open
System Model}
\author{Xinyu Zhao$^{1}$%
\footnote{Email:xzhao1@stevens.edu}%
, Wufu Shi$^{1}$%
\footnote{Email:wshi1@stevens.edu}%
, Lian-Ao Wu$^{2}$ and Ting Yu$^{1}$%
\footnote{Email:Ting.Yu@stevens.edu}%
}
\affiliation{$^{1}$Center for Controlled Quantum Systems and the Department of Physics and
Engineering Physics, Stevens Institute of Technology, Hoboken, New Jersey
07030, USA}
\affiliation{$^{2}$Department of Theoretical Physics and History of Science, The Basque
Country University (EHU/UPV), PO Box 644, 48080 Bilbao, Spain}

\begin{abstract}
This paper considers the extension of the non-Markovian stochastic approach
for quantum open systems strongly coupled to a fermionic bath, to the models
in which the system operators commute with the fermion bath. This technique
can also be a useful tool for studying open quantum systems coupled to a
spin-chain environment, which can be further transformed into an effective
fermionic bath. We derive an exact stochastic Schr\"{o}dinger equation (SSE),
called fermionic quantum state diffusion (QSD) equation, from the first
principle by using the fermionic coherent state representation. The reduced
density operator for the open system can be recovered from the average of the
solutions to the QSD equation over the Grassmann-type noise. By employing the
exact fermionic QSD equation, we can derive the corresponding exact master
equation. The power of our approach is illustrated by the applications of our
stochastic approach to several models of interest including the one-qubit
dissipative model, the coupled two-qubit dissipative model, the quantum
Brownian motion model and the N-fermion model coupled to a fermionic bath.
Different effects caused by the fermionic and bosonic baths on the dynamics of
open systems are also discussed.

\end{abstract}

\maketitle

\section{Introduction}

The theory of open quantum systems has experienced a resurgent interest
because of the rapid development of quantum experimental technologies and
their applications to the fabrication and manipulation of quantum devices
(e.g. photonic devices, quantum dots, nano-mechanical oscillators). However,
an intricate problem exists since in reality no system can be completely
isolated from its environment (bath, reservoir etc), and the dynamics of the
system of interest will be profoundly affected by the couplings to its
environment \cite{Gardiner1,Breuer}. When the quantum open systems are coupled
to a Markov environment, the Lindblad master equation is a critical tool which
can be used to study the dynamics of the open systems \cite{Lindblad}. When
the Born-Markov approximation is no longer valid, namely, the coupling between
system and environment is not weak and the environment cannot be approximated
by a broadband bath, one must extend the standard Markov theory to a more
general non-Markovian environment. Several attempts to derive the evolution
equation of open quantum systems beyond the Markov approximation have been
proposed \cite{Breuer,H-P-Z,Halliwell96,ZhangWM-2cav,ZhangWM-Tu,Sabrina}.
Notably, the non-Markovian quantum state diffusion (QSD) approach developed by
Strunz and his coworkers has showed momentous potential of solving large
systems (multi-qubit or multi-cavity)
\cite{QSD,Yu1999,YuQBM,Jing-Yu2010,JunArxiv,Xinyu2011,Broadbent2011}.
Moreover, as a computing tool, many numerical advantages of the QSD approach
permit its use in several domains such as high-precision measurement
\cite{Chen}, entanglement dynamics \cite{JingEPL} and coherence dynamics of
the large molecules in biophysics \cite{Eisfeld2011} etc. Therefore, it is
highly desirable to extend the QSD approach for the bosonic baths to the
fermionic case where the non-Markovian features have played increasingly
important role \cite{Search2002,Sinha2010,1qu-in-BF,1quFgas,QBM-Fbath}.

The primary theme of our current paper is to establish an exact quantum
approach for a class of quantum systems interacting with a fermionic bath. We
will consider a class of systems such that the systems and fermionic bath are
distinguishable, hence the system Hamiltonian and the bath operators commute.
The system of interest in this case may consist of one or more effective
particles such as spins, effective fermions etc. The case where the system and
bath operators ant-commute will be covered in-depth in a separate paper
\cite{Wufu2011}. It is noted that the commutative model we proposed arises
from many physical settings including spin bath and fermionic bath (e.g. see
Appendix A).

We will derive a fermionic stochastic Schr\"{o}dinger equation for an open
quantum system embedded in a fermionic bath, called the fermionic QSD
equation. To illustrate the power of our approach, we solve several models as
examples by using this new technique, including a one-qubit dissipative model,
a two-qubit dissipative model, the quantum Brownian motion in a fermionic bath
and a multiple-particle model. In the first example, we give the explicit
analytical solution without any approximation in a special case. In the second
example, we show how to construct the crucial $\hat{Q}$ operator contained in
the fermionic QSD equation. In the third example, we consider a continuous
variable model where a Brownian particle is immersed in a bath of fermion
particles. The last example involves a genuine multi-particle system that has
been solved exactly by our QSD approach. Finally, the difference between the
bosonic bath and the fermonic bath is discussed.

This paper is organized as follows. In section~\ref{fqsd}, we introduce the
general commutative fermionic bath model and derive the fundamental dynamic
equation for this model. In section~\ref{sec3}, we derive the formal exact
master equation from the QSD equation. In section~\ref{sec4}, we present a
simple example of using this fermionic QSD approach to solve the one-qubit
dissipative model. In section~\ref{sec5}, we solve the two-qubit dissipative
model to show the construction of some complicated $\hat{Q}$ operator. In
section~\ref{sec6}, we apply our fermionic QSD approach to a continuous
variable model to solve the quantum Brownian motion model in a fermionic bath.
In section~\ref{sec7}, we solve a genuine multi-partite system, the N-fermion
model, to show that our approach is not only applicable to small systems
(one-qubit or two-qubit), but is also applicable to large quantum open
systems. Based on the last example, we also evaluate the differences between
the bosonic bath and the fermonic bath. Finally, in section~\ref{sec8}, we
conclude the paper. In Appendix A, we provide an effective commutative model
consisting of spinless fermions as an environment. In Appendix B and C, we
present the details of derivation of the non-Markovian QSD equation for a
fermionic bath. In Appendix D, we prove a Novikov-type theorem for a Grassmann
Gaussian stochastic process, which plays a crucial role in deriving the exact
master equation from the corresponding stochastic Schr\"{o}dinger equation. In
Appendix E-F, we derive explicit equations of motion for the coefficients of
the master equation for the examples presented in this paper.


\section{Non-Markovian QSD equation for an open system coupled to a fermionic
bath}

\label{fqsd}


For a quantum open system interacting with a fermionic environment, the total
Hamiltonian may be written as%
\begin{equation}
\hat{H}_{\mathrm{tot}}=\hat{H}_{\mathrm{s}}+\hat{H}_{\mathrm{b}}+\hat
{H}_{\mathrm{int}}, \label{H_fermion_tot}%
\end{equation}
where $\hat{H}_{\mathrm{s}}$ is the Hamiltonian of the system, $\hat
{H}_{\mathrm{b}}$ is the Hamiltonian of the bath and $\hat{H}_{\mathrm{int}}$
is the interaction term. When we consider the fermionic bath, $\hat
{H}_{\mathrm{b}}$ and $\hat{H}_{\mathrm{int}}$ can be written as (setting
$\hbar=1$ throughout the paper)%
\begin{equation}
\hat{H}_{\mathrm{b}}=%
{\displaystyle\sum\limits_{i}}
\omega_{i}\hat{c}_{i}^{\dag}\hat{c}_{i},
\end{equation}%
\begin{equation}
\hat{H}_{\mathrm{int}}=%
{\displaystyle\sum\limits_{i}}
(g_{i}^{\ast}\hat{c}_{i}^{\dag}\hat{L}+g_{i}\hat{L}^{\dag}\hat{c}_{i}),
\label{H_fermion_int}%
\end{equation}
where $\hat{c}_{i}^{\dag}$ and $\hat{c}_{i}$\ are fermionic creation and
annihilation operators $\{\hat{c}_{i},\hat{c}_{j}^{\dag}\}=\delta_{ij}$.
Here we emphasize that the bath may consist of a set fermions or spins (e.g.
see \cite{Barouch1970}; An example is shown in Appendix A.)

In the interaction picture, the total Hamiltonian becomes%
\begin{equation}
\hat{H}_{\mathrm{tot}}(t)=\hat{H}_{\mathrm{s}}+%
{\displaystyle\sum\limits_{i}}
(g_{i}^{\ast}e^{i\omega_{i}t}\hat{c}_{i}^{\dag}\hat{L}+g_{i}e^{-i\omega_{i}%
t}\hat{L}^{\dag}\hat{c}_{i}).
\end{equation}
We use the fermionic coherent state (e.g., see \cite{ZhangRMP,Glauber1999}) to
describe the state of environment. For a single mode, the fermionic coherent
state is defined as%
\begin{equation}
\hat{c}_{i}|\xi_{i}\rangle=\xi_{i}|\xi_{i}\rangle,
\end{equation}
where $\xi_{i}$ is a Grassmann variable which satisfies the following
properties $\{\xi_{i},\xi_{j}\}=0,$ $\{\xi_{i},\xi_{j}^{\ast}\}=0.$ Generally,
the coherent state can be expanded in terms of Fock states as $|\xi_{i}%
\rangle=|0\rangle-\xi_{i}c_{i}^{\dagger}|0\rangle$. The coherent states (ket
and bra vectors) for the multi-mode environment are given by $|\xi\rangle
=|\xi_{1}\rangle\otimes|\xi_{2}\rangle\otimes|\xi_{3}\rangle\otimes...$ and
$\langle\xi|=\langle\xi_{1}|\otimes\langle\xi_{2}|\otimes\langle\xi
_{3}|\otimes....$

Now, we can define%
\begin{equation}
\psi_{t}(\xi^{\ast})=\langle\xi|\psi_{\mathrm{tot}}(t)\rangle,
\end{equation}
where $|\psi_{\mathrm{tot}}(t)\rangle$ is the total state vector for the
system and environment, and $\langle\xi|$ is a coherent state representation
for the environment. In this paper, we focus on the case where the initial
state of the bath is vacuum state. The finite temperature bath will be
discussed in \cite{Wufu2011} by using the Bogoliubov transformation
\cite{Yu-FiniteT}. With the coherent state representation, we can derive the
non-Markovian QSD equation for the fermionic bath as%
\begin{equation}
\frac{\partial}{\partial t}\psi_{t}(\xi^{\ast})=[-i\hat{H}_{\mathrm{s}}%
+\hat{L}\xi_{t}^{\ast}-\hat{L}^{\dag}{\int\nolimits_{0}^{t}} dsK(t,s)\frac
{\delta_{l}}{\delta\xi_{s}^{\ast}}]\psi_{t}(\xi^{\ast}),
\end{equation}
where $\xi_{t}^{\ast}=-i{\sum\nolimits_{i}} g_{i}^{\ast}e^{i\omega_{i}t}%
\xi_{i}^{\ast}$ and $K(t,s)={\sum\nolimits_{i}} |g_{i}|^{2}e^{-i\omega
_{i}(t-s)}$ is the correlation function. (Details of the derivation can be
found in Appendix B) We use $\frac{\delta_{l}}{\delta\xi_{s}^{\ast}}\psi
_{t}(\xi^{\ast})$ to denote the left-functional-derivative with respect to the
Grassmann variables. Our fermionic approach is applicable to arbitrary
correlation functions especially for the general non-Markovian case.

Similar to the formal bosonic QSD equation \cite{QSD}, the fermionic QSD
contains a time-nonlocal Grassmann functional derivative which renders a
direct application of the derived fermionic QSD extremely difficult, if not
impossible. In order to find a time-local QSD equation, one can introduce a
time-dependent operator (also $\xi^{\ast}$-dependent in general) $\hat{Q}$,
defined as%
\begin{equation}
\frac{\delta_{l}\psi_{t}(\xi^{\ast})}{\delta\xi_{s}^{\ast}}=\hat{Q}%
(t,s,\xi^{\ast})\psi_{t}(\xi^{\ast}).
\end{equation}
If no confusion arise, we will use the shorthand notation: $\hat{Q}=\hat
{Q}(t,s,\xi^{\ast})$. With this $\hat{Q}$ operator, the exact stochastic QSD
equation can be written as%
\begin{equation}
\frac{\partial}{\partial t}\psi_{t}(\xi^{\ast})=[-i\hat{H}_{\mathrm{s}}%
+\hat{L}\xi_{t}^{\ast}-\hat{L}^{\dag}\bar{Q}]\psi_{t}(\xi^{\ast}), \label{QSD}%
\end{equation}
where $\bar{Q}(t,\xi^{\ast})={\int\nolimits_{0}^{t}}dsK(t,s)\hat{Q}%
(t,s,\xi^{\ast})$. The stochastic QSD equation for a fermionic bath we have
presented here is an exact equation of motion for the open quantum system
directly derived from the microscopic Hamiltonian without any approximation.
It should be noted that in our derivation of the QSD equation, we have not
explicitly specified the system Hamiltonian and the coupling operators,
$\hat{H}_{\mathrm{s}}$ and $\hat{L}$. Here we have introduced a new type of
stochastic process $\xi_{t}^{\ast}$. The solution of our QSD equation is
called a Grassmann quantum trajectory. By construction, the reduced density
matrix of the open system can be recovered by the statistical mean over the
Grassmann noise. Although the fermionic QSD equation looks formally similar to
the bosonic case, the dynamic behaviors of the system governed by the two
types of equations can be different due to distinct differences between the
bosonic and fermionic particles. Mathematically, the most striking difference
between the bosonic and fermionic QSD equations is that the former contains a
complex Gaussian noise while the latter is driven by a non-commutative
Grassmann Gausian noise.We will illustrate the difference in a concrete
example in a subsequent section.

In order to derive the dynamic equation for the $\hat{Q}$ operator, we
consider the consistency condition (CC),
\begin{equation}
\frac{\delta_{l}}{\delta\xi_{s}^{\ast}}\frac{\partial}{\partial t}\psi_{t}%
(\xi^{\ast})=\frac{\partial}{\partial t}\frac{\delta_{l}}{\delta\xi_{s}^{\ast
}}\psi_{t}(\xi^{\ast}).
\end{equation}
Applying the QSD Eq. (\ref{QSD}) to CC, the equation for $\hat{Q}$ operator is
derived as%
\begin{align}
\frac{\partial}{\partial t}\hat{Q}  &  =-i[\hat{H}_{\mathrm{s}},\hat
{Q}]-\{\hat{L}\xi_{t}^{\ast},\hat{Q}\}\nonumber\\
&  -\hat{L}^{\dag}\bar{Q}(-\xi^{\ast})\hat{Q}+\hat{Q}\hat{L}^{\dag}\bar
{Q}-\hat{L}^{\dag}\frac{\delta}{\delta\xi_{s}^{\ast}}\bar{Q}, \label{Eq_O}%
\end{align}
where the sign of $\bar{Q}(-\xi^{\ast})$ depends on the functional form of
noise contained in $\bar{Q}$. (Details of derivation and discussion can be
found in Appendix C.) The initial condition for the $\hat{Q}$ operator is%
\begin{equation}
\hat{Q}(t,s=t,\xi^{\ast})=\hat{L}.
\end{equation}
However, for a simple case that $\hat{Q}$ is independent of Grassmann noise,
the equation for the $\hat{Q}$ reduces to%
\begin{equation}
\frac{\partial}{\partial t}\hat{Q}=-i[\hat{H}_{\mathrm{s}},\hat{Q}]-\{\hat
{L}\xi_{t}^{\ast},\hat{Q}\}-[\hat{L}^{\dag}\bar{Q},\hat{Q}]. \label{Eq_O_free}%
\end{equation}
Eq. (\ref{Eq_O}) and Eq. (\ref{Eq_O_free}) can be used to determine the exact
$\hat{Q}$ operator. However, for most practical problems, it may be a daunting
task to determine the exact $\hat{Q}$. Therefore, it is important to develop a
perturbation approach similar to that developed for the bosonic bath
\cite{Yu1999}. In fact, we may expand $\hat{Q}$ operator as%
\begin{align}
\hat{Q}(t,s,\xi^{\ast})  &  =\hat{Q}^{(0)}(t,s)+\int_{0}^{t}\hat{Q}%
^{(1)}(t,s,s_{1})\xi_{s_{1}}^{\ast}ds_{1}\nonumber\\
&  +\int_{0}^{t}\int_{0}^{t}\hat{Q}^{(2)}(t,s,s_{1},s_{2})\xi_{s_{1}}^{\ast
}\xi_{s_{2}}^{\ast}ds_{1}ds_{2}+...\nonumber\\
&  +\int_{0}^{t}...\int_{0}^{t}\hat{Q}^{(n)}(t,s,s_{1},...s_{n})\xi_{s_{1}%
}^{\ast}...\xi_{s_{n}}^{\ast}ds_{1}...ds_{n}\nonumber\\
&  +....
\end{align}
Substituting this equation into Eq.~(\ref{Eq_O}), one can derive the dynamic
equations of the coefficients for each order $\hat{Q}^{(i)}$. Particularly,
the zeroth-order term $\hat{Q}^{(0)}(t,s)$ will satisfy the following equation
(neglect all the noise terms)%
\begin{equation}
\frac{\partial}{\partial t}\hat{Q}^{(0)}(t,s)=-i[\hat{H}_{\mathrm{s}},\hat
{Q}^{(0)}(t,s)]-[\hat{L}^{\dag}\bar{Q}^{(0)}(t),\hat{Q}^{(0)}(t,s)],
\label{Eq_O_zero}%
\end{equation}
where $\bar{Q}^{(0)}(t)=\int_{0}^{t}\hat{Q}^{(0)}(t,s)K(t,s)ds$, and the
initial condition is%
\begin{equation}
\hat{Q}^{(0)}(t,s=t)=\hat{L}.
\end{equation}

\section{Formal exact master equation for an open quantum system coupled to a
fermionic bath}

\label{sec3}

Now, we will derive the master equation governing the reduced density operator
of the open quantum system from the stochastic QSD equation (\ref{QSD}).
First, we define the stochastic density operator as
\begin{equation}
\hat{P}_{t}=\left\vert \psi_{t}(\xi^{\ast})\right\rangle \left\langle \psi
_{t}(-\xi)\right\vert
\end{equation}
It is easy to verify that the reduced density matrix of the open system can be
written as
\begin{align}
\hat{\rho}  &  ={\sum\limits_{n}}\langle n|\psi_{\mathrm{tot}}\rangle
\langle\psi_{\mathrm{tot}}|n\rangle\nonumber\\
&  =\int%
{\displaystyle\prod\nolimits_{i}}
d\xi_{i}^{\ast}d\xi_{i}e^{-\sum_{j}\xi_{j}^{\ast}\xi_{j}}%
{\displaystyle\sum\limits_{n}}
\langle n|\xi\rangle\langle\xi|\psi_{\mathrm{tot}}\rangle\langle
\psi_{\mathrm{tot}}|n\rangle\nonumber\\
&  =\int%
{\displaystyle\prod\nolimits_{i}}
d\xi_{i}^{\ast}d\xi_{i}e^{-\sum_{j}\xi_{j}^{\ast}\xi_{j}}%
{\displaystyle\sum\limits_{n}}
\langle\xi|\psi_{\mathrm{tot}}\rangle\langle\psi_{\mathrm{tot}}|n\rangle
\langle n|-\xi\rangle\nonumber\\
&  =\int%
{\displaystyle\prod\nolimits_{i}}
d\xi_{i}^{\ast}d\xi_{i}e^{-\sum_{j}\xi_{j}^{\ast}\xi_{j}}\hat{P}%
_{t}\nonumber\\
&  =\langle\hat{P}_{t}\rangle_{s}%
\end{align}
where $\langle...\rangle_{s}$ denotes the statistical mean over the Grassmann
Gaussian noise defined by%
\begin{equation}
\langle...\rangle_{s}\equiv\int{\prod\nolimits_{i}}d\xi_{i}^{\ast}d\xi
_{i}e^{-\sum_{j}\xi_{j}^{\ast}\xi_{j}}(...).
\end{equation}
From this expression, we say that the reduced density matrix can be unraveled
by a set of Grassmann quantum trajectories $|\psi_{t}(\xi^{\ast})\rangle$.

From the QSD equation Eq.~(\ref{QSD}), we have%
\begin{equation}
\frac{\partial}{\partial t}\left\langle \psi_{t}(-\xi)\right\vert
=\left\langle \psi_{t}(-\xi)\right\vert [i\hat{H}_{\mathrm{s}}-\xi_{t}\hat
{L}^{\dagger}-\bar{Q}^{\dagger}(t,-\xi)\hat{L}],
\end{equation}
thus,%
\begin{align}
\frac{\partial}{\partial t}\hat{\rho}  &  =\frac{\partial}{\partial t}%
\langle\hat{P}_{t}\rangle_{s}\nonumber\\
&  =\langle(-i\hat{H}_{\mathrm{s}}+\hat{L}\xi_{t}^{\ast}-\hat{L}^{\dag}\bar
{Q})\hat{P}_{t}\rangle_{s}\nonumber\\
&  +\langle\hat{P}_{t}(i\hat{H}_{\mathrm{s}}-\xi_{t}\hat{L}^{\dagger}-\bar
{Q}^{\dagger}(-\xi)\hat{L})\rangle_{s}\nonumber\\
&  =-i[\hat{H}_{\mathrm{s}},\hat{\rho}]+\hat{L}\langle\xi_{t}^{\ast}\hat
{P}_{t}\rangle_{s}-\langle\hat{P}_{t}\xi_{t}\rangle_{s}\hat{L}^{\dagger
}\nonumber\\
&  -\hat{L}^{\dag}\langle\bar{Q}\hat{P}_{t}\rangle_{s}-\langle\hat{P}_{t}%
\bar{Q}^{\dagger}(-\xi)\rangle_{s}\hat{L}.
\end{align}

In order to establish the exact master equation from the fermionic QSD
equation (\ref{QSD}), one needs to handle the terms $\langle\hat{P}_{t}\xi
_{t}\rangle_{s}$ etc. In fact, we can prove a Novikov-type theorem for the
Grassmann Gaussian noise (see Appendix D),%
\begin{align}
\langle\hat{P}_{t}\xi_{t}\rangle_{s}  &  =-\langle\bar{Q}\hat{P}_{t}%
\rangle_{s},\\
\langle\xi_{t}^{\ast}\hat{P}_{t}\rangle_{s}  &  =\langle\hat{P}_{t}\bar
{Q}^{\dagger}(-\xi)\rangle_{s}.
\end{align}
With the help of the Novikov-type theorem for the Grassmann noise, the exact
master equation can be written as%
\begin{equation}
\frac{\partial}{\partial t}\hat{\rho}=-i[\hat{H}_{\mathrm{s}},\hat{\rho
}]+[\hat{L},\langle\hat{P}_{t}\bar{Q}^{\dagger}(-\xi)\rangle_{s}]+[\langle
\bar{Q}\hat{P}_{t}\rangle_{s},\hat{L}^{\dagger}].
\end{equation}
If the operator $\hat{Q}$ is independent of the Grassmann noise, then the
exact master equation is immediately obtained,%
\begin{equation}
\frac{\partial}{\partial t}\hat{\rho}=-i[\hat{H}_{\mathrm{s}},\hat{\rho
}]+[\hat{L},\hat{\rho}\bar{Q}^{\dagger}]+[\bar{Q}\hat{\rho},\hat{L}^{\dagger
}].
\end{equation}
Moreover, in the Markov limit, $\bar{Q}=\gamma_{f}\hat{L}$, this master
equation reduces to the standard Lindblad master equation:
\begin{equation}
\frac{\partial}{\partial t}\hat{\rho}=-i[\hat{H}_{\mathrm{s}},\hat{\rho
}]+\gamma_{f}[\hat{L},\hat{\rho}\hat{L}^{\dagger}]+\gamma_{f}[\hat{L}\hat
{\rho},\hat{L}^{\dagger}].
\end{equation}
In subsequent sections, we will derive several interesting master equations
from the corresponding QSD equations.

\section{Example 1: One-qubit dissipative model}

\label{sec4}

We start with a very simple example, one-qubit in fermionic bath. This is a
special case where it is possible to derive the fully analytical solution
without any approximation.

\subsection{Master equation and non-Markovian quantum dynamics}

The total Hamiltonian for the one-qubit dissipative model may be written as
\begin{equation}
\hat{H}_{\mathrm{tot}}=\hat{H}_{\mathrm{s}}+\hat{H}_{\mathrm{b}}+\hat
{H}_{\mathrm{int}},
\end{equation}%
\begin{equation}
\hat{H}_{\mathrm{s}}=\frac{\omega}{2}\hat{\sigma}_{z},
\end{equation}%
\begin{equation}
\hat{H}_{\mathrm{b}}={\sum\limits_{i}} \omega_{i}\hat{c}_{i}^{\dag}\hat{c}%
_{i},
\end{equation}%
\begin{equation}
\hat{H}_{\mathrm{int}}={\sum\limits_{i}} (g_{i}^{\ast}\hat{c}_{i}^{\dag}%
\hat{L}+g_{i}\hat{L}^{\dag}\hat{c}_{i}),
\end{equation}
where $\hat{L}=\hat{\sigma}_{-}$ for this particular model.

From Eq. (\ref{Eq_O_free}), the solution for the $\hat{Q}$ can be obtained as%
\begin{equation}
\hat{Q}(t,s)=x_{1}(t,s)\hat{\sigma}_{-}, \label{ansatz_of_O}%
\end{equation}
with the initial condition%
\begin{equation}
\hat{Q}(t,s=t)=\hat{L}=\hat{\sigma}_{-},
\end{equation}
and the coefficient $x_{1}(t,s)$ is shown to satisfy%
\begin{equation}
\frac{\partial}{\partial t}x_{1}(t,s)=[i\omega+X_{1}(t)]x_{1}(t,s),
\end{equation}
where $X_{1}(t)={\int\nolimits_{0}^{t}} x_{1}(t,s)K(t,s)ds$, and $K(t,s)$ is
the correlation function, and the initial condition is given by $x_{1}%
(t,s)=1.$

Thus, the exact $\hat{Q}$ operator can be fully determined. It is worth noting
that this $\hat{Q}$ operator has the same form as the bosonic case
\cite{Yu1999}. Finally, the explicit QSD equation for this model is%
\begin{equation}
\frac{\partial}{\partial t}\psi_{t}(\xi^{\ast})=[-i\frac{\omega}{2}\hat
{\sigma}_{z}+\hat{\sigma}_{-}\xi_{t}^{\ast}-X_{1}(t)\hat{\sigma}_{+}%
\hat{\sigma}_{-}]\psi_{t}(\xi^{\ast}),
\end{equation}
and the exact master equation is%
\begin{align}
\frac{d}{dt}\hat{\rho}  &  =-i[\hat{H}_{\mathrm{s}},\hat{\rho}]+[\hat{L}%
,\hat{\rho}\hat{Q}^{\dagger}]+[\hat{Q}\hat{\rho},\hat{L}^{\dagger}]\nonumber\\
&  =-i\frac{\omega}{2}(\hat{\sigma}_{z}\hat{\rho}-\hat{\rho}\hat{\sigma}%
_{z})+X_{1}^{\ast}(t)(\hat{\sigma}_{-}\hat{\rho}\hat{\sigma}_{+}-\hat{\rho
}\hat{\sigma}_{+}\hat{\sigma}_{-})\nonumber\\
&  +X_{1}(t)(\hat{\sigma}_{-}\hat{\rho}\hat{\sigma}_{+}-\hat{\sigma}_{+}%
\hat{\sigma}_{-}\hat{\rho}). \label{MEQ_1qu}%
\end{align}
With this exact master equation, the dynamics of this model can be fully determined.


\subsection{A limiting case -- the environment consists of only one fermion}

Now, we consider a very special case for the one-qubit model where the
\textquotedblleft environment" \cite{comment} contains only one fermion. By
analytically solving this model, we show explicitly that the fermionic QSD
gives rise to identical results to those predicted by the ordinary quantum
mechanics. The model is described by the following Hamiltonian,%
\begin{equation}
\hat{H}_{\mathrm{tot}}=\frac{\omega}{2}\hat{\sigma}_{z}+\omega_{b}\hat
{c}^{\dagger}\hat{c}+(g^{\ast}\hat{\sigma}_{-}\hat{c}^{\dagger}+g\hat{\sigma
}_{+}\hat{c}), \label{Hamiltonian_of_1mode}%
\end{equation}
and the zero-temperature correlation function becomes%
\begin{equation}
K(t,s)=\left\vert g\right\vert ^{2}e^{-i\omega_{b}(t-s)}.
\end{equation}
Substituting the correlation function into the expression of $X_{1}%
(t)={\int\nolimits_{0}^{t}} x_{1}(t,s)K(t,s)ds$, we will find the differential
equation for $X_{1}(t)$ as%
\begin{equation}
\frac{\partial}{\partial t}X_{1}(t)=\left\vert g\right\vert ^{2}-i\omega
_{b}X_{1}(t)+i\omega X_{1}(t)+X_{1}(t)^{2}.
\end{equation}

For simplicity, we consider the resonance case, then the solution $X_{1}(t)$
can reduce to%
\begin{equation}
X_{1}(t)=\left\vert g\right\vert \tan(\left\vert g\right\vert t).
\end{equation}

From the master equation Eq. (\ref{MEQ_1qu}), we can calculate time evolution
for the off-diagonal elements in the density matrix.%
\begin{equation}
\frac{d}{dt}\hat{\rho}_{21}=\frac{d}{dt}\left\langle \hat{\sigma}%
_{+}\right\rangle =Tr(\frac{d}{dt}\hat{\rho}\hat{\sigma}_{+})=i\omega\hat
{\rho}_{21}-X_{1}^{\ast}(t)\hat{\rho}_{21}.
\end{equation}
Finally, we can derive the time evolution for $\hat{\rho}_{21}$ as
\begin{equation}
\hat{\rho}_{21}(t)=\hat{\rho}_{21}(0)e^{i\omega t}\cos\left[  \left\vert
g\right\vert t\right]  . \label{Solution_QSD}%
\end{equation}
Similarly, we can get%
\begin{equation}
\hat{\rho}_{12}(t)=\hat{\rho}_{12}(0)e^{-i\omega t}\cos\left[  \left\vert
g\right\vert t\right]  . \label{Solution_QSD2}%
\end{equation}
This result shows that the coherence (off-diagonal elements in density matrix)
will decrease and increase periodically.

On the other hand, we can easily solve this simple case using elementary
quantum mechanics. Since this is only a two-body problem, we can solve the
evolution for the whole system in a straightforward manner. One can check that
elementary quantum mechanics gives rise to the identical results obtained by
the fermionic QSD approach in Eq.~(\ref{Solution_QSD},\ref{Solution_QSD2}).

\section{Example 2: Coupled two-qubit dissipative model}

\label{sec5} In this section, we consider a system containing a pair of
coupled two-level systems (spins or some other effective two-level models)
interacting with a common fermionic bath. We will show how to construct exact
and approximate $\hat{Q}$ operator in this example. The total Hamiltonian of
this model can be written as,
\begin{equation}
\hat{H}_{\mathrm{tot}}=\hat{H}_{\mathrm{s}}+\hat{H}_{\mathrm{b}}+\hat
{H}_{\mathrm{int}}, \label{Hamiltonian}%
\end{equation}
where%
\begin{align}
\hat{H}_{\mathrm{s}}  &  =\omega_{A}\hat{\sigma}_{z}^{A}+\omega_{B}\hat
{\sigma}_{z}^{B}+J_{xy}(\hat{\sigma}_{+}^{A}\hat{\sigma}_{-}^{B}+\hat{\sigma
}_{-}^{A}\hat{\sigma}_{+}^{B})+J_{z}\hat{\sigma}_{z}^{A}\hat{\sigma}_{z}%
^{B},\nonumber\\
\hat{H}_{\mathrm{b}}  &  =\sum_{j}\omega_{j}\hat{c}_{j}^{\dagger}\hat{c}%
_{j},\nonumber\\
\hat{H}_{\mathrm{int}}  &  =\sum_{j}(g_{j}\hat{c}_{j}^{\dagger}\hat{L}%
+g_{j}^{\ast}\hat{c}_{j}\hat{L}^{\dagger}),
\end{align}
Here, the operator $\hat{L}=\kappa_{A}\hat{\sigma}_{-}^{A}+\kappa_{B}%
\hat{\sigma}_{-}^{B}$ describles the pattern of interaction to the
environment. $\kappa_{A}$ and $\kappa_{B}$ are constants describing different
coupling strengths for the two qubits.

The perturbative zeroth-order $\hat{Q}$ operator can be derived as%
\begin{equation}
\hat{Q}^{(0)}(t,s)=\sum_{i=1}^{4}f_{i}(t,s)\hat{Q}_{i}%
\end{equation}
where $\hat{Q}_{i}$ ($i=1,2,3,4$) are the time-independent basis operators,
and $f_{i}(t,s)$ are time-dependent coefficients.

The four basis operators in terms of the Pauli matrices may be written as%
\begin{equation}
\hat{Q}_{1}=\hat{\sigma}_{-}^{A},\,\,\hat{Q}_{2}=\hat{\sigma}_{-}^{B}%
,\,\,\hat{Q}_{3}=\hat{\sigma}_{z}^{A}\hat{\sigma}_{-}^{B},\text{ \ }\hat
{Q}_{4}=\hat{\sigma}_{z}^{B}\hat{\sigma}_{-}^{A},
\end{equation}
From Eq. (\ref{Eq_O_zero}), we can derive the differential equation for the
coefficients as \begin{widetext}
\begin{align}
\frac{\partial}{\partial t}f_{1}(t,s)  &  =+2i\omega_{A}f_{1}-iJ_{xy}%
f_{3}+2iJ_{z}f_{4}+\kappa_{A}F_{1}f_{1}-\kappa_{B}F_{1}f_{3}+\kappa_{B}%
F_{3}f_{1}+\kappa_{B}F_{3}f_{4}+\kappa_{B}F_{4}f_{3}+\kappa_{A}F_{4}f_{4},\\
\frac{\partial}{\partial t}f_{2}(t,s)  &  =+2i\omega_{B}f_{2}-iJ_{xy}%
f_{4}+2iJ_{z}f_{3}+\kappa_{B}F_{2}f_{2}-\kappa_{A}F_{2}f_{4}+\kappa_{B}%
F_{3}f_{3}+\kappa_{A}F_{3}f_{4}+\kappa_{A}F_{4}f_{2}+\kappa_{A}F_{4}f_{3},\\
\frac{\partial}{\partial t}f_{3}(t,s)  &  =+2i\omega_{B}f_{3}-iJ_{xy}%
f_{1}+2iJ_{z}f_{2}-\kappa_{A}F_{2}f_{1}+\kappa_{B}F_{2}f_{3}+\kappa_{A}%
F_{3}f_{1}+\kappa_{A}F_{4}f_{2}+\kappa_{B}F_{3}f_{2}+\kappa_{A}F_{4}f_{3},\\
\frac{\partial}{\partial t}f_{4}(t,s)  &  =+2i\omega_{A}f_{4}-iJ_{xy}%
f_{2}+2iJ_{z}f_{1}-\kappa_{B}F_{1}f_{2}+\kappa_{A}F_{1}f_{4}+\kappa_{B}%
F_{3}f_{1}+\kappa_{B}F_{3}f_{4}+\kappa_{A}F_{4}f_{1}+\kappa_{B}F_{4}f_{2},
\end{align}%
\end{widetext}where $F_{i}(t)=\int_{0}^{t}dsK(t,s)f_{i}(t,s)$ ($i=1,2,3,4$),
and the initial conditions are
\begin{align}
f_{1}(t,s  &  =t)=\kappa_{A},\\
f_{2}(t,s  &  =t)=\kappa_{B},\\
f_{3}(t,s  &  =t)=0,\\
f_{4}(t,s  &  =t)=0,
\end{align}

Moreover, we can also determine the exact $\hat{Q}$ operator for this
two-qubit model. We can verify rigorously that the exact $\hat{Q}$ operator
contains five terms where the last term is noise-dependent. The details of the
derivation is presented in Appendix E. If we use this zeroth-order $\hat{Q}$
operator, the master equation can be explicitly written in the following form%
\begin{align}
\frac{d}{dt}\hat{\rho}  &  =-i[\hat{H}_{\mathrm{s}}\hat{\rho}-\hat{\rho}%
\hat{H}_{\mathrm{s}}]\nonumber\\
&  +\{\sum_{i=1}^{4}F_{i}^{\ast}[\hat{L}\hat{\rho}\bar{Q}_{i}^{\dagger}%
-\hat{\rho}\bar{Q}_{i}^{\dagger}\hat{L}]+H.C.\}.
\end{align}
Next, we consider a simple case, in which all the parameters are symmetric for
two qubits, i.e. $\omega_{A}=\omega_{B}=\omega,$ $\kappa_{A}=\kappa_{B}=1.$
Then, we can derive the following master equation%
\begin{align}
\frac{d}{dt}\hat{\rho}  &  =-i[\hat{H}_{\mathrm{s}},\hat{\rho}]+[\hat{L}%
,\hat{\rho}\bar{Q}^{\dagger}]+[\bar{Q}\hat{\rho},\hat{L}^{\dagger}]\nonumber\\
&  =-i\omega\lbrack(\hat{\sigma}_{z}^{A}+\hat{\sigma}_{z}^{B})\hat{\rho}%
-\hat{\rho}(\hat{\sigma}_{z}^{A}+\hat{\sigma}_{z}^{B})]\nonumber\\
&  -iJ_{xy}[(\hat{\sigma}_{+}^{A}\hat{\sigma}_{-}^{B}+\hat{\sigma}_{+}^{B}%
\hat{\sigma}_{-}^{A})\hat{\rho}-\hat{\rho}(\hat{\sigma}_{+}^{A}\hat{\sigma
}_{-}^{B}+\hat{\sigma}_{+}^{B}\hat{\sigma}_{-}^{A})]\nonumber\\
&  -iJ_{z}[\hat{\sigma}_{z}^{A}\hat{\sigma}_{z}^{B}\hat{\rho}-\hat{\rho}%
\hat{\sigma}_{z}^{A}\hat{\sigma}_{z}^{B}]\nonumber\\
&  +\{F_{1}^{\ast}[(\hat{\sigma}_{-}^{A}+\hat{\sigma}_{-}^{B})\hat{\rho}%
\hat{\sigma}_{+}^{A}-\hat{\rho}\hat{\sigma}_{+}^{A}(\hat{\sigma}_{-}^{A}%
+\hat{\sigma}_{-}^{B})]\nonumber\\
&  +F_{2}^{\ast}[(\hat{\sigma}_{-}^{A}+\hat{\sigma}_{-}^{B})\hat{\rho}%
\hat{\sigma}_{+}^{B}-\hat{\rho}\hat{\sigma}_{+}^{B}(\hat{\sigma}_{-}^{A}%
+\hat{\sigma}_{-}^{B})]\nonumber\\
&  +F_{3}^{\ast}[(\hat{\sigma}_{-}^{A}+\hat{\sigma}_{-}^{B})\hat{\rho}%
\hat{\sigma}_{z}^{A}\hat{\sigma}_{+}^{B}-\hat{\rho}\hat{\sigma}_{z}^{A}%
\hat{\sigma}_{+}^{B}(\hat{\sigma}_{-}^{A}+\hat{\sigma}_{-}^{B})]\nonumber\\
&  +F_{4}^{\ast}[(\hat{\sigma}_{-}^{A}+\hat{\sigma}_{-}^{B})\hat{\rho}%
\hat{\sigma}_{z}^{B}\hat{\sigma}_{+}^{A}-\hat{\rho}\hat{\sigma}_{z}^{B}%
\hat{\sigma}_{+}^{A}(\hat{\sigma}_{-}^{A}+\hat{\sigma}_{-}^{B})]+H.C.\}.
\end{align}

The master equation derived above is valid for a general correlation function.
For numerical simulations, one need to consider a special example of the
correlation function. A general correlation function may be written as%
\begin{align}
K(t,s)  &  =\int_{0}^{\infty}d\omega J(\omega)[\coth(\omega/2k_{B}T)\cos
\omega(t-s)\nonumber\\
&  -i\sin\omega(t-s)],
\end{align}
where $J(\omega)$ is the spectral density. If we choose $J(\omega
)=\Gamma\omega e^{-(\frac{\omega}{_{\omega_{c}}})}$, which is so-called Ohmic
case, the correlation function in the zero-temperature can be written as%
\begin{equation}
K(t,s)=\frac{\Gamma}{[\frac{1}{\omega_{c}}+i(t-s)]^{2}},
\end{equation}
where $\omega_{c}$ is the cut-off frequency.

\begin{figure}[ptb]
\begin{center}
\includegraphics[
trim=0.000000in 0.000000in 0.000000in -0.801684in,
height=3.0113in,
width=3.5129in
]{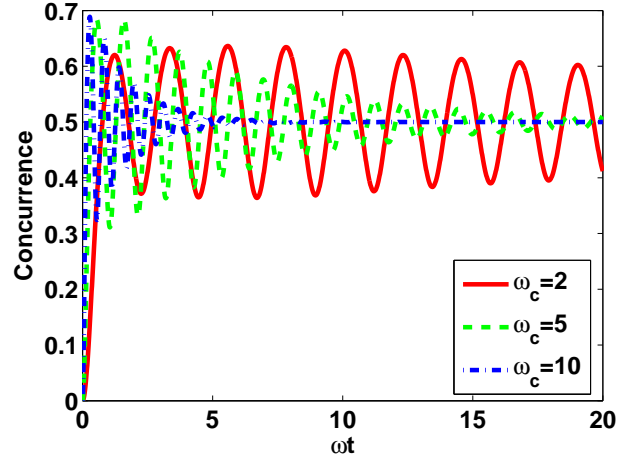}
\end{center}
\caption{Time evolution of concurrence for different $\omega_{c}$.}%
\label{c_wc}%
\end{figure}

\begin{figure}[ptb]
\begin{center}
\includegraphics[
trim=0.000000in 0.000000in 0.000000in -0.540555in,
height=2.0081in,
width=3.2119in
]{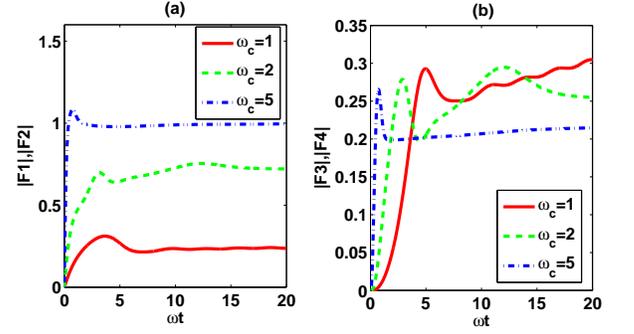}
\end{center}
\caption{Time evolution of $|F_{1}(t)|$, $|F_{2}(t)|$, $|F_{3}(t)|$, and
$|F_{4}(t)|$. In the symmetric case, $|F_{1}(t)|=|F_{2}(t)|$, $|F_{3}%
(t)|=|F_{4}(t)|$. The parameters are $\omega_{1}=\omega_{2}=\omega=1$,}%
\label{F1234}%
\end{figure}

\section{Example 3: Quantum Brownian particle in a fermionic bath}

\label{sec6}

We consider a continuous model consisting of a Brownian particle interacting
with a fermionic bath. The Hamiltonian of the Brownian particle is given by,
\begin{equation}
\hat{H}_{\mathrm{s}}=\omega_{m}(\hat{p}^{2}+\hat{q}^{2}).
\end{equation}
The Hamiltonian of the fermionic bath is%
\begin{equation}
\hat{H}_{\mathrm{b}}={\sum\limits_{i}} \omega_{i}\hat{c}_{i}^{\dagger}\hat
{c}_{i},
\end{equation}
and the interaction Hamiltonian is given by%
\begin{equation}
\hat{H}_{\mathrm{int}}=\hat{q}{\sum\limits_{i}} (g_{i}^{\ast}\hat{c}%
_{i}^{\dagger}+g_{i}\hat{c}_{i}).
\end{equation}
So, the total Hamiltonian is%
\begin{equation}
\hat{H}_{\mathrm{tot}}=\hat{H}_{\mathrm{s}}+\hat{H}_{\mathrm{b}}+\hat
{H}_{\mathrm{int}}.
\end{equation}
Applying our QSD approach to this model, it can be easily shown that $\hat{Q}$
operator takes the following form:%
\begin{align}
\hat{Q}  &  =x_{1}(t,s)\hat{q}+x_{2}(t,s)\hat{p}+x_{3}(t,s,\xi^{\ast})\hat
{p}\hat{q}\nonumber\\
&  +x_{4}(t,s,\xi^{\ast})\hat{p}^{2}+x_{5}(t,s,\xi^{\ast})\hat{q}^{2}+....
\end{align}
which is a infinite series, therefore, it is difficult to determine the exact
$\hat{Q}$ operator. A useful approximation is to neglect all the
noise-dependent terms, after which we obtain the so-called zeroth-order
approximate $\hat{Q}$ as%
\begin{equation}
\hat{Q}\approx x_{1}(t,s)\hat{q}+x_{2}(t,s)\hat{p}.
\end{equation}
Substituting this approximate $\hat{Q}$ operator into Eq. (\ref{Eq_O_zero}),
we can derive the differential equations for the coefficients $x_{1}(t,s)$ and
$x_{2}(t,s)$ as%
\begin{align}
\frac{\partial}{\partial t}x_{1}(t,s)  &  =2\omega_{m}x_{2}(t,s)+iX_{2}%
(t)x_{1}(t,s)-2iX_{1}(t)x_{2}(t,s),\\
\frac{\partial}{\partial t}x_{2}(t,s)  &  =-2\omega_{m}x_{1}(t,s)-iX_{2}%
(t)x_{2}(t,s).
\end{align}
The initial conditions for coefficients $x_{1}(t,s)$ and $x_{2}(t,s)$ are%
\begin{align}
x_{1}(t,s  &  =t)=1,\\
x_{2}(t,s  &  =t)=0.
\end{align}
Using this approximate $\hat{Q}$ operator, the master equation can be written
as%
\begin{align}
\frac{d}{dt}\hat{\rho}  &  =-i[\hat{H}_{\mathrm{s}},\hat{\rho}]+[\hat{L}%
,\hat{\rho}\bar{Q}^{\dagger}]+[\bar{Q}\hat{\rho},\hat{L}^{\dagger}]\nonumber\\
&  =-i\omega_{m}[(\hat{p}^{2}+\hat{q}^{2})\hat{\rho}-\hat{\rho}(\hat{p}%
^{2}+\hat{q}^{2})]\nonumber\\
&  +\{X_{1}^{\ast}[\hat{q}\hat{\rho}\hat{q}-\hat{\rho}\hat{q}\hat{q}%
]+X_{2}^{\ast}[\hat{q}\hat{\rho}\hat{p}-\hat{\rho}\hat{p}\hat{q}]+H.C.\}.
\end{align}
It should be noted that the $X_{2}^{\ast}$ (including its conjugation $X_{2}$)
does not exist in the Markov limit. Hence, the approximate $\hat{Q}$ operator
defined above is different from the Markov approximation. It is also different
from the weak-coupling approximation since the approximate $\hat{Q}$ still
contains the higher-order terms of the coupling constant. We expect that the
master equation obtained from the approximate $\hat{Q}$ will be valid in a
weakly non-Markovian regime.

From the master equation we can derive the evolution equations for all the
mean values of operators $\hat{q},\hat{p}$,
\begin{align}
\frac{d}{dt}\langle\hat{q}\rangle &  =2\omega_{m}\langle\hat{p}\rangle,\\
\frac{d}{dt}\langle\hat{p}\rangle &  =-2\omega_{m}\langle\hat{q}\rangle
-iX_{1}^{\ast}\langle\hat{q}\rangle-iX_{2}^{\ast}\langle\hat{p}\rangle
+iX_{1}\langle\hat{q}\rangle+iX_{2}\langle\hat{p}\rangle.
\end{align}
\begin{figure}[ptb]
\begin{center}
\includegraphics[
trim=0.000000in 0.000000in 0.000000in -0.166965in,
height=2.5097in,
width=3.2119in
]{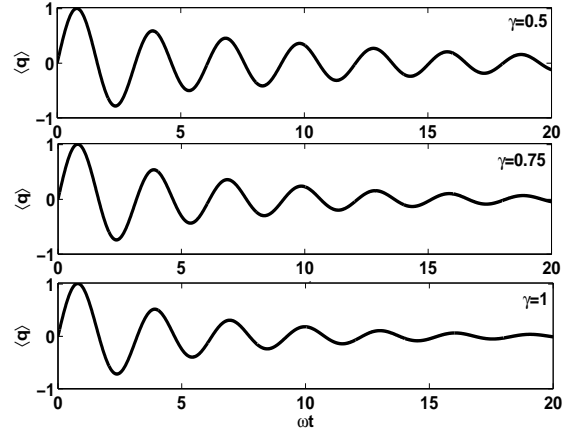}
\end{center}
\caption{Time evolution of mean values of operator $\hat{q}$ in different
environments. The parameter $\gamma$ indicates the memory effect. The other
parameters are $\omega_{m}=\omega=1,$ $\Omega=\pi/2$.}%
\label{meanq.eps}%
\end{figure}

In Fig.~\ref{meanq.eps}, we plot the time evolution of $\langle\hat{q}\rangle$
in different kinds of environments with different $\gamma$. In order to show
the transition from non-Markovian to Markovian, Ornstein-Uhlenbeck noise
$K(t,s)=\frac{\gamma}{2}e^{-(\gamma+i\Omega)|t-s|}$ is chosen in our numerical
simulations. The reason of using Ornstein-Uhlenbeck noise is that the memory
time of the environment can be described by one parameter $1/\gamma$.
Fig.~\ref{meanq.eps} shows how the evolution of $\langle\hat{q}\rangle$ is
affected by $\gamma$. This is a unique phenomenon in the non-Markovian case.

\section{Example 4. N-fermion system coupled to a fermionic bath}

\label{sec7}

\subsection{Dynamic equation for the general N-fermion model}

In the last example, we will establish the exact time-local fermionic QSD
equation and master equation for a genuine multipartite system coupled to a
fermionic bath. We show that, using the fermionic QSD approach, the exact
$\hat{Q}$ operator of the N-qubit model can be easily determined.

More specifically, let us consider the following Hamiltonian%
\begin{align}
\hat{H}_{\mathrm{tot}}  &  =\hat{H}_{\mathrm{s}}+\hat{H}_{\mathrm{b}}+\hat
{H}_{\mathrm{int}},\label{Nfermion_tot}\\
\hat{H}_{\mathrm{b}}  &  ={\sum\limits_{j=1}^{N_{b}}} \omega_{j}\hat{c}%
_{j}^{\dagger}\hat{c}_{j},\label{Nfermion_bath}\\
\hat{H}_{\mathrm{int}}  &  ={\sum\limits_{j=1}^{N_{b}}} g_{j}(\hat{c}%
_{j}^{\dagger}\hat{L}+\hat{L}^{\dagger}\hat{c}_{j}). \label{Nfermion_int}%
\end{align}
where $\hat{H}_{\mathrm{b}}$\ is a fermionic bath. We assume that the system
of interest consists of $N$ fermions, i.e.%
\begin{equation}
\hat{H}_{\mathrm{s}}=\sum_{i=1}^{N_{s}}A_{i}\hat{a}_{i}^{\dagger}\hat{a}_{i},
\end{equation}
here, $\hat{a}_{i}^{\dagger}$\ and $\hat{a}_{i}$\ are also fermion creation
and annihilation operators; the Lindblad operator is%
\begin{equation}
\hat{L}=\sum_{i=1}^{N_{s}}\hat{a}_{i}. \label{Nfermion_L}%
\end{equation}
This Hamiltonian could be an effective Hamiltonian transformed from a set of
spins. For example, suppose that we have a long chain with $N$ sites, if the
first $N_{s}(N_{s}<N)$ sites are treated as system and the other $N_{b}$ sites
are treated as bath ($N_{s}+N_{b}=N$), then performing the Jordan-Wigner
transformations for both the system and the bath, we may result in this type
of effective Hamiltonian (for details, see Appendix A).

We can show that the exact $\hat{Q}$ operator of this model takes the
following form%
\begin{equation}
\hat{Q}=\sum_{i=1}^{N_{s}}x_{i}(t,s)\hat{a}_{i},
\end{equation}
and the differential equation for the coefficients in $\hat{Q}$ operator are
given by%
\begin{equation}
\frac{\partial}{\partial t}x_{j}(t,s)=iA_{j}x_{j}(t,s)+\sum_{i}^{N_{s}}%
X_{j}(t)x_{i}(t,s), \label{OF_coeff}%
\end{equation}
where $X_{j}(t)=\int_{0}^{t}K(t,s)x_{j}(t,s)ds$. So, $\bar{Q}(t)=\sum
_{i=1}^{N_{s}}X_{i}(t)\hat{a}_{i}$. The exact master equation of this model is%
\begin{align}
\frac{\partial}{\partial t}\hat{\rho}  &  =-i[\hat{H}_{\mathrm{s}},\hat{\rho
}]+[\hat{L},\hat{\rho}\bar{Q}^{\dagger}]+[\bar{Q}\hat{\rho},\hat{L}^{\dagger
}]\nonumber\\
&  =-i[\hat{H}_{\mathrm{s}},\hat{\rho}]+[\sum_{i=1}^{N_{s}}\hat{a}_{i}%
,\hat{\rho}\sum_{i}^{N_{s}}X_{i}^{\ast}(t)\hat{a}^{\dagger}]\nonumber\\
&  +[(\sum_{i=1}^{N_{s}}X_{i}(t)\hat{a}_{i})\hat{\rho},\sum_{i=1}^{N_{s}}%
\hat{a}_{i}^{\dagger}].
\end{align}

\subsection{Fermionic versus bosonic baths}

It is instructive to consider a simple case with two fermions in the system
($N_{s}=2$). The Hamiltonian is then given by
\begin{align}
\hat{H}_{\mathrm{s}}  &  =\omega_{1}\hat{a}_{1}^{\dagger}\hat{a}_{1}%
+\omega_{2}\hat{a}_{2}^{\dagger}\hat{a}_{2},\\
\hat{L}  &  =\hat{a}_{1}+\hat{a}_{2},
\end{align}
it is easy to show that the exact $\bar{Q}$ operator is
\begin{equation}
\bar{Q}=X_{1}(t)\hat{a}_{1}+X_{2}(t)\hat{a}_{2} \label{FO}%
\end{equation}
where $X_{1}(t,s)$ and $X_{2}(t,s)$ can be determined in Eq. (\ref{OF_coeff})
as $N_{s}=2$ case. Then, the explicit master equation can be written as%
\begin{align}
\frac{d}{dt}\hat{\rho}  &  =-i\omega_{1}(\hat{a}_{1}^{\dagger}\hat{a}_{1}%
\hat{\rho}-\hat{\rho}\hat{a}_{1}^{\dagger}\hat{a}_{1})-i\omega_{2}(\hat{a}%
_{2}^{\dagger}\hat{a}_{2}\hat{\rho}-\hat{\rho}\hat{a}_{2}^{\dagger}\hat{a}%
_{2})\nonumber\\
&  +\{X_{1}^{\ast}(t)(\hat{a}_{1}\hat{\rho}\hat{a}_{1}^{\dagger}-\hat{\rho
}\hat{a}_{1}^{\dagger}\hat{a}_{1})+X_{1}^{\ast}(t)(\hat{a}_{2}\hat{\rho}%
\hat{a}_{1}^{\dagger}-\hat{\rho}\hat{a}_{1}^{\dagger}\hat{a}_{2})\nonumber\\
&  +X_{2}^{\ast}(t)(\hat{a}_{1}\hat{\rho}\hat{a}_{2}^{\dagger}-\hat{\rho}%
\hat{a}_{2}^{\dagger}\hat{a}_{1})+X_{2}^{\ast}(t)(\hat{a}_{2}\hat{\rho}\hat
{a}_{2}^{\dagger}-\hat{\rho}\hat{a}_{2}^{\dagger}\hat{a}_{2})\nonumber\\
&  +H.C.\}.
\end{align}

On the other hand, we can also solve this model exactly if the two effective
fermions (spins) are coupled to a bosonic bath. The Hamiltonian takes the same
form as Eq. (\ref{Nfermion_tot}-\ref{Nfermion_L}), except that $\hat{c}_{j}$
($\hat{c}_{j}^{\dagger}$) represent bosonic annihilation (creation) operators
(also consider $N_{s}=2$ case). Using the non-Markovian QSD approach for
bosonic bath \cite{QSD}, the bosonic QSD equation can be derived as%
\begin{equation}
\frac{\partial}{\partial t}\psi_{t}(z^{\ast})=[-i\hat{H}_{\mathrm{s}}+\hat
{L}z_{t}^{\ast}-\hat{L}^{\dag}\bar{O}]\psi_{t}(z^{\ast}),
\end{equation}
where $\bar{O}(t,z^{\ast})={\int\nolimits_{0}^{t}}dsK(t,s)\hat{O}(t,s,z^{\ast
})$. In the bosonic QSD equation, the noise $z_{t}^{\ast}=-i{\sum
\nolimits_{i}}g_{i}^{\ast}e^{i\omega_{i}t}z_{i}^{\ast}$ is the complex (not
Grassmann) Gaussian noise.
The exact $\bar{O}$ operator is determined as follows%
\begin{align}
\bar{O}(t,z^{\ast})  &  =X_{1}(t)\hat{a}_{1}+X_{2}(t)\hat{a}_{2}+X_{3}%
(t)\hat{a}_{1}^{\dagger}\hat{a}_{1}\hat{a}_{2}\nonumber\\
&  +X_{4}(t)\hat{a}_{2}^{\dagger}\hat{a}_{1}\hat{a}_{2}+i\int_{0}%
^{t}ds^{\prime}X_{5}(t,s^{\prime})z_{s^{\prime}}^{\ast}\hat{a}_{1}\hat{a}_{2}.
\label{BO}%
\end{align}
Details about the coefficients can be found in Appendix F.

\begin{figure}[ptb]
\begin{center}
\includegraphics[
trim=0.000000in 0.000000in 0.000000in -0.148155in,
height=2.5097in,
width=3.5129in
]{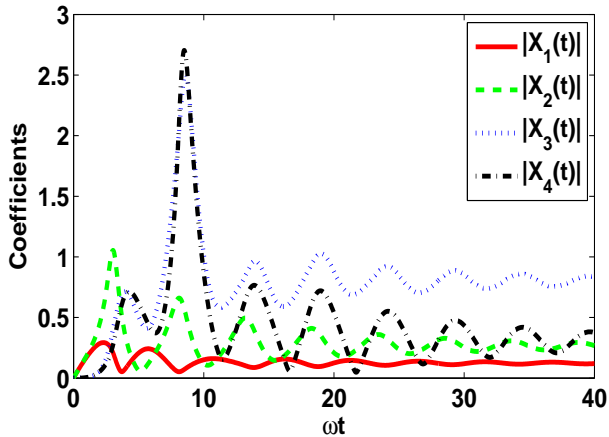}
\end{center}
\caption{Dynamic evolution of the coefficients in $\bar{Q}$ (for fermionic
bath) and $\bar{O}$ (for bosonic bath)\ operators. The parameters are
$\omega_{1}=2$, $\omega_{2}=\omega=1$, $\gamma=0.4$, $\Omega=\pi/4$.}%
\label{coeff}%
\end{figure}

We use this particular example to illustrate different aspects between the
fermionic and bosonic baths. As shown above, we can find the exact $\hat{Q}$
($\hat{O}$) operators for both the fermnionic bath and the bosonic
counterpart. Since the exact dynamic evolution of the system will be fully
determined by $\hat{Q}$ ($\hat{O}$) operators, so we may compare the
difference between the two operators given in Eq.~(\ref{FO}) and Eq.
(\ref{BO}), respectively. The first two terms $X_{1}(t,s)$ and $X_{2}(t,s)$
are the same for both the $\hat{Q}$ and $\hat{O}$ operators (one can easily
check that they satisfy the same equations), and the difference comes from
other terms. In the cases where $X_{1}(t,s)$ and $X_{2}(t,s)$ are dominant,
one would not expects sharp difference between the fermionic and bosonic
baths. For example, when $\omega_{1}=\omega_{2}$, two operators $\hat{Q}$ and
$\hat{O}$ are exactly the same. However, we found that the extra terms
$X_{3}(t,s)$ and $X_{4}(t,s)$ occurred in $\hat{O}$ may become important under
certain conditions as shown in Fig.~\ref{coeff} where the coefficients in the
$\hat{O}$ and $\hat{Q}$ operators are plotted. Clearly, the fermionic and
bosonic baths may result in very different dynamics. In the numerical
simulations, we choose the Ornstein-Uhlenbeck noise $K(t,s)=\frac{\gamma}%
{2}e^{-(\gamma+i\Omega)|t-s|}$ for simplicity. However, our approach is
applicable for arbitrary kinds of correlation function.

\section{Conclusion}

\label{sec8}

In summarizing, we have developed a novel technique called the fermionic
quantum state diffusion approach which is a useful tool for studying quantum
open systems coupled to a fermionic bath. Using the Grassmann coherent state,
the exact fermionic QSD equation and the corresponding master equation are
derived for several physically interesting models. We have shown that the
time-local QSD approach developed in this paper can efficiently solve open
systems coupled to fermionic baths by employing the exact or approximate
$\hat{Q}$ operators. Moreover, our research also suggests that some spin bath
problems can also be solved by using the effective fermion bath. Finally, it
is of great interest to apply the fermionic QSD approach to more realistic
models such as finite temperature fermion baths and large spin baths, and we
leave these topics open for future discussion.


\section*{Acknowledgements}

We thank Prof. J. H. Eberly, Prof. N. J. M. Horing, Prof. B. L. Hu, and Dr. J.
Jing for useful discussions and the support by grants from the NSF
PHY-0925174, AFOSR No. FA9550-12-1-0001, Ikerbasque Foundation Startup, the
Basque Government (grant IT472-10) and the Spanish MEC (Project No. FIS2009-12773-C02-02).

\appendix

\section{Spin chain as an effective fermion bath model}

In this section, we consider a spin-chain model where some spins are treated
as the system of interest, the rest is treated as its environment. We show
that the model can be transformed to a fermionic bath model.

Consider a quantum system interacting with a XX spin chain. The Hamiltonian is%
\begin{align}
\hat{H}_{\mathrm{tot}}  &  =\hat{H}_{\mathrm{s}}+\hat{H}_{\mathrm{b}}+\hat
{H}_{\mathrm{int}},\label{H_spin_tot}\\
\hat{H}_{\mathrm{b}}  &  =\sum_{i}(\hat{\sigma}_{i}^{+}\hat{\sigma}_{i+1}%
^{-}+\hat{\sigma}_{i+1}^{+}\hat{\sigma}_{i}^{-}),\\
\hat{H}_{\mathrm{int}}  &  =\hat{L}^{\dagger}\hat{\sigma}_{1}+\hat{\sigma}%
_{1}^{\dagger}\hat{L}. \label{H_spin_int}%
\end{align}

After performing the Jordan-Wigner transformation,%
\begin{equation}
\hat{\sigma}_{j}^{-}=\exp(-i\pi\sum_{k=1}^{j-1}\hat{c}_{k}^{\dagger}\hat
{c}_{k})\hat{c}_{j},
\end{equation}
and the Fourier transformation \cite{Barouch1970},%
\begin{equation}
\hat{c}_{j}=\frac{1}{\sqrt{N}}\sum_{p=-N/2}^{N/2}\exp(-ij\phi_{p})\hat{a}_{p},
\end{equation}
the original Hamiltonian Eq. (\ref{H_spin_tot}-\ref{H_spin_int}) become%
\begin{equation}
\hat{H}_{\mathrm{tot}}=\hat{H}_{\mathrm{s}}+\hat{H}_{\mathrm{b}}+\hat
{H}_{\mathrm{int}},
\end{equation}%
\begin{equation}
\hat{H}_{\mathrm{b}}=\sum_{p=-N/2}^{N/2}2\cos\phi_{p}\hat{a}_{p}^{\dagger}%
\hat{a}_{p},
\end{equation}%
\begin{equation}
\hat{H}_{\mathrm{int}}=\frac{1}{\sqrt{N}}\sum_{p=-N/2}^{N/2}[\hat{L}^{\dagger
}\exp(-i\phi_{p})\hat{a}_{p}+\exp(i\phi_{p})\hat{a}_{p}^{\dagger}\hat{L}].
\end{equation}
This effective Hamiltonian obtained from the transformation takes the same
form given by Eq. (\ref{H_fermion_tot}-\ref{H_fermion_int}). Therefore, we may
use the QSD approach to study the dynamics of the subsystem of the spin-chain model.

\section{Derivation of the non-Markovian QSD equation for a fermionic bath}

To start with, we list several useful commutation relations between fermionic
coherent state and operators:
\begin{align}
\langle\xi_{i}|\hat{L}  &  =\hat{L}\langle\xi_{i}|,\text{ }\langle\xi_{i}%
|\hat{H}_{\mathrm{s}}=\hat{H}_{\mathrm{s}}\langle\xi_{i}|,\nonumber\\
\langle\xi_{i}|\hat{c}_{i}  &  =\frac{\partial_{l}}{\partial\xi_{i}^{\ast}%
}\langle\xi_{i}|,\text{ }\langle\xi_{i}|\hat{c}_{i}^{\dagger}=\langle\xi
_{i}|\xi_{i}^{\ast}=\xi_{i}^{\ast}\langle\xi_{i}|.
\end{align}
Using these relations, we can derive the QSD equation as%
\begin{align}
&  \frac{\partial}{\partial t}\psi_{t}(\xi^{\ast})\nonumber\\
&  =-i\langle\xi|\hat{H}_{\mathrm{tot}}(t)|\psi_{\mathrm{tot}}(t)\rangle
\nonumber\\
&  =-i\langle\xi|\hat{H}_{\mathrm{s}}+%
{\displaystyle\sum\limits_{i}}
(g_{i}^{\ast}e^{i\omega_{i}t}\hat{c}_{i}^{\dag}\hat{L}+H.C.)|\psi
_{\mathrm{tot}}(t)\rangle\nonumber\\
&  =-i\hat{H}_{\mathrm{s}}\psi_{t}(\xi^{\ast})+\hat{L}\xi_{t}^{\ast}\psi
_{t}(\xi^{\ast})\nonumber\\
&  -i\hat{L}^{\dag}%
{\displaystyle\sum\limits_{i}}
g_{i}e^{-i\omega_{i}t}\langle\xi|\hat{c}_{i}|\psi_{\mathrm{tot}}(t)\rangle,
\end{align}
where $\xi_{t}^{\ast}=-i{\sum\nolimits_{i}} g_{i}^{\ast}e^{i\omega_{i}t}%
\xi_{i}^{\ast}$. Then using the chain rule to introduce the functional
derivative,%
\begin{align}
&  \langle\xi|\hat{c}_{i}|\psi_{\mathrm{tot}}(t)\rangle=\frac{\partial_{l}%
}{\partial\xi_{i}^{\ast}}\psi_{t}(\xi^{\ast})\\
&  =\int ds\frac{\partial\xi_{s}^{\ast}}{\partial\xi_{i}^{\ast}}\frac
{\delta_{l}}{\delta\xi_{s}^{\ast}}\psi_{t}(\xi^{\ast}).
\end{align}
Finally, we have
\begin{equation}
\frac{\partial}{\partial t}\psi_{t}(\xi^{\ast})=[-i\hat{H}_{\mathrm{s}}%
+\hat{L}\xi_{t}^{\ast}-\hat{L}^{\dag}\int dsK(t,s)\frac{\delta_{l}}{\delta
\xi_{s}^{\ast}}]\psi_{t}(\xi^{\ast}),
\end{equation}
where $K(t,s)={\sum\limits_{i}} |g_{i}|^{2}e^{-i\omega_{i}(t-s)}$. This is
just the final QSD equation.

\section{Equation for $\hat{Q}$ operator}

First, consider the following two commutation relations:%
\begin{equation}
\frac{\delta_{l}}{\delta\xi_{s}^{\ast}}[\xi_{t}^{\ast}\psi_{t}(\xi^{\ast
})]=-\xi_{t}^{\ast}\frac{\delta_{l}}{\delta\xi_{s}^{\ast}}\psi_{t}(\xi^{\ast})
\label{dfztpsi}%
\end{equation}
and%
\begin{equation}
\frac{\delta_{l}}{\delta\xi_{s}^{\ast}}[\bar{Q}\psi_{t}(\xi^{\ast})]=\bar
{Q}(-\xi^{\ast})\hat{Q}\psi_{t}(\xi^{\ast})+(\frac{\delta_{l}}{\delta\xi
_{s}^{\ast}}\bar{Q})\psi_{t}(\xi^{\ast}), \label{dfopsi}%
\end{equation}
for fixed order of $\hat{Q}$ operator. One can prove them easily.

With Eq. (\ref{dfztpsi}) and Eq. (\ref{dfopsi}), we can apply the consistency
condition to $\psi_{t}(\xi^{\ast})$.%
\begin{equation}
\frac{\delta_{l}}{\delta\xi_{s}^{\ast}}\frac{\partial}{\partial t}\psi_{t}%
(\xi^{\ast})=\frac{\partial}{\partial t}\frac{\delta_{l}}{\delta\xi_{s}^{\ast
}}\psi_{t}(\xi^{\ast}).
\end{equation}
The left-hand side is%
\begin{align}
LHS  &  =[-i\hat{H}_{\mathrm{s}}\hat{Q}-\hat{L}\xi_{t}^{\ast}\hat
{Q}\nonumber\\
&  -\hat{L}^{\dag}(\frac{\delta}{\delta\xi_{s}^{\ast}}\bar{Q})-\hat{L}^{\dag
}\bar{Q}(-\xi^{\ast})\hat{Q}]\psi_{t}(\xi^{\ast}). \label{LHS}%
\end{align}

On the other hand, the right-hand side becomes%
\begin{equation}
RHS=\frac{\partial}{\partial t}(\hat{Q})\psi_{t}(\xi^{\ast})+[-i\hat{Q}\hat
{H}_{\mathrm{s}}+\hat{Q}\hat{L}\xi_{t}^{\ast}-\hat{Q}\hat{L}^{\dag}\bar
{Q}]\psi_{t}(\xi^{\ast}). \label{RHS}%
\end{equation}
Equate LHS and RHS and eliminate $\psi_{t}(\xi^{\ast})$, we have%
\begin{align}
\frac{\partial}{\partial t}\hat{Q}  &  =-i[\hat{H}_{\mathrm{s}},\hat
{Q}]-\{\hat{L}\xi_{t}^{\ast},\hat{Q}\}\nonumber\\
&  -\hat{L}^{\dag}\bar{Q}(-\xi^{\ast})\hat{Q}+\hat{Q}\hat{L}^{\dag}\bar
{Q}-\hat{L}^{\dag}\frac{\delta_{l}}{\delta\xi_{s}^{\ast}}\bar{Q}.
\end{align}

\section{Proof of Novikov-type theorem for the Grassmann noise}

In this section, we will provide a proof of Novikov-type theorem for a
Grassmann Gaussian noise, which plays a crucial role in deriving the exact or
approximate master equations from the corresponding stochastic Schr\"{o}dinger equations.

\emph{Theorem}: Suppose that $\xi_{t},\xi_{t}^{\ast}$ are Grassmann-type
Gaussian processes and the $\hat{P}_{t}$ is the stochastic density operator,
then we have the following two identities:
\begin{align}
\langle\hat{P}_{t}\xi_{t}\rangle_{s}  &  =-\langle\bar{Q}(\xi^{\ast})\hat
{P}_{t}\rangle_{s},\\
\langle\xi_{t}^{\ast}\hat{P}_{t}\rangle_{s}  &  =\langle\hat{P}_{t}\bar
{Q}^{\dagger}(-\xi)\rangle_{s}.
\end{align}

\emph{Proof:}
\begin{widetext}%
\begin{align}
&  \langle\hat{P}_{t}\xi_{t}\rangle_{s}\nonumber\\
&  =%
{\displaystyle\int}
{\displaystyle\prod\nolimits_{i}}
d\xi_{i}^{\ast}d\xi_{i}e^{-\sum_{i}\xi_{i}^{\ast}\xi_{i}}\left\vert \psi
(\xi^{\ast})\right\rangle \left\langle \psi(-\xi)\right\vert (i\sum_{j}%
g_{j}e^{-i\omega_{j}t}\xi_{j})\nonumber\\
&  =-i\sum_{j}g_{j}e^{-i\omega_{j}t}%
{\displaystyle\int}
{\displaystyle\prod\nolimits_{i}}
d\xi_{i}^{\ast}d\xi_{i}[\left\vert \psi(\xi^{\ast})\right\rangle \left\langle
\psi(-\xi)\right\vert \frac{\partial_{l}}{\partial\xi_{j}^{\ast}}(e^{-\sum
_{i}\xi_{i}^{\ast}\xi_{i}})]\nonumber\\
&  =i\sum_{j}g_{j}e^{-i\omega_{j}t}%
{\displaystyle\int}
{\displaystyle\prod\nolimits_{i}}
d\xi_{i}^{\ast}d\xi_{i}(\frac{\partial_{l}}{\partial(-\xi_{j}^{\ast}%
)}\left\vert \psi(\xi^{\ast})\right\rangle \left\langle \psi(-\xi)\right\vert
)e^{-\sum_{i}\xi_{i}^{\ast}\xi_{i}}\nonumber\\
&  =-i\sum_{j}g_{j}e^{-i\omega_{j}t}%
{\displaystyle\int}
{\displaystyle\prod\nolimits_{i}}
d\xi_{i}^{\ast}d\xi_{i}[e^{-\sum_{i}\xi_{i}^{\ast}\xi_{i}}(\int ds\frac
{\partial\xi_{s}^{\ast}}{\partial\xi_{j}^{\ast}}\frac{\delta_{l}}{\delta
\xi_{s}^{\ast}})\hat{P}_{t}]\nonumber\\
&  =-\int ds\sum_{j}\left\vert g_{j}\right\vert ^{2}e^{-i\omega_{j}(t-s)}%
{\displaystyle\int}
{\displaystyle\prod\nolimits_{i}}
d\xi_{i}^{\ast}d\xi_{i}[e^{-\sum_{i}\xi_{i}^{\ast}\xi_{i}}\frac{\delta_{l}%
}{\delta\xi_{s}^{\ast}}\hat{P}_{t}]\nonumber\\
&  =-%
{\displaystyle\int}
{\displaystyle\prod\nolimits_{i}}
d\xi_{i}^{\ast}d\xi_{i}e^{-\sum_{i}\xi_{i}^{\ast}\xi_{i}}\int dsK(t,s)\hat
{Q}(t,s,\xi^{\ast})\hat{P}_{t}]\nonumber\\
&  =-\langle\bar{Q}\hat{P}_{t}\rangle_{s}.
\end{align}%
\end{widetext}%
Similarly, we can prove%
\begin{equation}
\langle\xi_{t}^{\ast}\hat{P}_{t}\rangle_{s}=\langle\hat{P}_{t}\bar{Q}%
^{\dagger}(-\xi)\rangle_{s}.
\end{equation}
This concludes our proof of the Novikov-type theorem for the Grassmann
Gaussian noise.

\section{Exact $\hat{Q}$ operator for the two-qubit model}

For the coupled two-qubit model, the exact $\hat{Q}$ takes the following form:%
\begin{align}
\hat{Q}(t,s,\xi^{\ast})  &  =f_{1}(t,s)\hat{Q}_{1}+f_{2}(t,s)\hat{Q}_{2}%
+f_{3}(t,s)\hat{Q}_{3}\nonumber\\
&  +f_{4}(t,s)\hat{Q}_{4}+i\int_{0}^{t}ds^{\prime}f_{5}(t,s,s^{\prime}%
)\xi_{s^{\prime}}^{\ast}\hat{Q}_{5}, \label{Ansztz of O}%
\end{align}
where the basis operators are given by
\begin{align}
\hat{Q}_{1}  &  =\hat{\sigma}_{-}^{A},\,\,\hat{Q}_{2}=\hat{\sigma}_{-}%
^{B},\,\,\hat{Q}_{3}=\hat{\sigma}_{z}^{A}\hat{\sigma}_{-}^{B},\nonumber\\
\hat{Q}_{4}  &  =\hat{\sigma}_{z}^{B}\hat{\sigma}_{-}^{A},\,\,\hat{Q}%
_{5}=2\hat{\sigma}_{-}^{A}\hat{\sigma}_{-}^{B},
\end{align}
and $f_{j}$ $(j=1,2,3,4,5)$ are some time-dependent coefficients. Substituting
Eq. (\ref{Ansztz of O}) into Eq. (\ref{Eq_O}), we obtain a set of partial
differential equations governing the coefficients of the $\hat{Q}$ operator,%
\begin{widetext}%
%

\begin{align}
\frac{\partial}{\partial t}f_{1}(t,s)  &  =+2i\omega_{A}f_{1}+\kappa_{A}%
F_{1}f_{1}+\kappa_{B}F_{3}f_{1}-iJ_{xy}f_{3}-\kappa_{B}F_{1}f_{3}+\kappa
_{B}F_{4}f_{3}+2iJ_{z}f_{4}+\kappa_{A}F_{4}f_{4}+\kappa_{B}F_{3}f_{4}%
-i\kappa_{B}F_{5},\nonumber\\
\frac{\partial}{\partial t}f_{2}(t,s)  &  =+2i\omega_{B}f_{2}+\kappa_{A}%
F_{4}f_{2}+\kappa_{B}F_{2}f_{2}+2iJ_{z}f_{3}+\kappa_{A}F_{4}f_{3}+\kappa
_{B}F_{3}f_{3}-iJ_{xy}f_{4}-\kappa_{A}F_{2}f_{4}+\kappa_{A}F_{3}f_{4}%
-i\kappa_{A}F_{5},\nonumber\\
\frac{\partial}{\partial t}f_{3}(t,s)  &  =-iJ_{xy}f_{1}-\kappa_{A}F_{2}%
f_{1}+\kappa_{A}F_{3}f_{1}+2iJ_{z}f_{2}+\kappa_{A}F_{4}f_{2}+\kappa_{B}%
F_{3}f_{2}+2i\omega_{B}f_{3}+\kappa_{A}F_{4}f_{3}+\kappa_{B}F_{2}f_{3}%
-i\kappa_{A}F_{5},\nonumber\\
\frac{\partial}{\partial t}f_{4}(t,s)  &  =+2iJ_{z}f_{1}+\kappa_{A}F_{4}%
f_{1}+\kappa_{B}F_{3}f_{1}-iJ_{xy}f_{2}-\kappa_{B}F_{1}f_{2}+\kappa_{B}%
F_{4}f_{2}+2i\omega_{A}f_{4}+\kappa_{A}F_{1}f_{4}+\kappa_{B}F_{3}f_{4}%
-i\kappa_{B}F_{5},\nonumber\\
\frac{\partial}{\partial t}f_{5}(t,s,s^{\prime})  &  =+\kappa_{A}F_{5}%
f_{1}+\kappa_{B}F_{5}f_{2}-\kappa_{B}F_{5}f_{3}-\kappa_{A}F_{5}f_{4}%
+2i\omega_{A}f_{5}+2i\omega_{B}f_{5}+\kappa_{A}F_{1}f_{5}+\kappa_{A}F_{4}%
f_{5}+\kappa_{B}F_{2}f_{5}+\kappa_{B}F_{3}f_{5},
\end{align}
%

\end{widetext}%
where $F_{j}(t)=\int_{0}^{t}dsK(t,s)f_{j}(t,s)$ $(j=1,2,3,4)$ and
$F_{5}(t,s^{\prime})=\int_{0}^{t}dsK(t,s)f_{5}(t,s,s^{\prime})$, with the
initial conditions:
\begin{align}
f_{1}(t,s  &  =t)=\kappa_{A},\nonumber\\
f_{2}(t,s  &  =t)=\kappa_{B},\nonumber\\
f_{3}(t,s  &  =t)=0,\nonumber\\
f_{4}(t,s  &  =t)=0,\nonumber\\
f_{5}(t,s  &  =t,s^{\prime})=0,\nonumber\\
f_{5}(t,s,s^{\prime}  &  =t)=i[\kappa_{A}f_{2}(t,s)+\kappa_{B}f_{1}(t,s)].
\end{align}


\section{Differential equations for coefficients of bosonic $\hat{O}$ in
example 4}

The coefficients in Eq. (\ref{BO}) satisfy the following differential
equations%
\begin{equation}
\frac{\partial}{\partial t}x_{1}(t,s)=i\omega_{a}x_{1}+x_{1}X_{1}+x_{2}X_{1},
\end{equation}%
\begin{equation}
\frac{\partial}{\partial t}x_{2}(t,s)=i\omega_{b}x_{2}+x_{1}X_{2}+x_{2}X_{2},
\end{equation}%
\begin{align}
\frac{\partial}{\partial t}x_{3}(t,s)  &  =i\omega_{b}x_{3}-x_{4}X_{2}%
+x_{3}X_{2}+x_{2}X_{3}\nonumber\\
&  +x_{3}X_{3}-x_{3}X_{4}-x_{2}X_{4}-iX_{5},
\end{align}%
\begin{align}
\frac{\partial}{\partial t}x_{4}(t,s)  &  =i\omega_{a}x_{4}+x_{4}X_{1}%
+x_{1}X_{4}-x_{1}X_{3}\nonumber\\
&  -x_{3}X_{1}+x_{4}X_{3}-x_{4}X_{4}-iX_{5},
\end{align}%
\begin{align}
\frac{\partial}{\partial t}x_{5}(t,s,s^{\prime})  &  =i\omega_{a}x_{5}%
+i\omega_{b}x_{5}+x_{5}X_{1}+x_{5}X_{2}\nonumber\\
&  +x_{5}X_{3}-x_{5}X_{4}+x_{1}X_{5}+x_{2}X_{5},
\end{align}
with the initial conditions%
\begin{align}
x_{1}(t,s  &  =t)=1,\\
x_{2}(t,s  &  =t)=1,\\
x_{3}(t,s  &  =t)=0,\\
x_{4}(t,s  &  =t)=0,\\
x_{5}(t,s  &  =t,s^{\prime})=0,\\
ix_{5}(t,s,s^{\prime}  &  =t)=2(x_{2}-x_{1})+x_{3}+x_{4}.
\end{align}
and%
\begin{align}
X_{j}(t)  &  =\int_{0}^{t}K(t,s)x_{j}(t,s)ds\text{ (}j=1\text{ to }4\text{)}\\
X_{5}(t,s^{\prime})  &  =\int_{0}^{t}K(t,s)x_{5}(t,s,s^{\prime})ds
\end{align}


\begin{thebibliography}{99}                                                                                               %


\bibitem {Gardiner1}C. W. Gardiner and P. Zoller, \emph{Quantum Noise}
(Springer-Verlag, Berlin, 2004).

\bibitem {Breuer}H. P. Breuer and F. Petruccione, \emph{Theory of Open Quantum
Systems} (Oxford University, New York, 2002).

\bibitem {Lindblad}G. Lindblad, Comm. Math. Phys. \textbf{48}, 119 (1976).

\bibitem {H-P-Z}B. L. Hu, J. P. Paz, and Y. Zhang, Phys. Rev. D \textbf{45},
2843 (1992), Phys. Rev. D \textbf{47}, 1567 (1993).

\bibitem {Halliwell96}J. J. Halliwell and T. Yu, Phys. Rev. D \textbf{53},
2012 (1996).

\bibitem {ZhangWM-2cav}J.-H. An and W.-M. Zhang, Phys. Rev. A \textbf{76},
042127 (2007).

\bibitem {ZhangWM-Tu}M. W. Y. Tu and W. -M. Zhang, Phys. Rev. B, \textbf{78},
235311 (2008).

\bibitem {Sabrina}S. Maniscalco and F. Petruccione, Phys. Rev. A. \textbf{73},
012111 (2006).





\bibitem {QSD}L. Di\'{o}si, N. Gisin, and W. T. Strunz, Phys. Rev. A
\textbf{58}, 1699 (1998); see also W. T. Strunz, L. Di\'{o}si, and N. Gisin,
Phys. Rev. Lett. \textbf{82}, 1801 (1999).

\bibitem {Yu1999}T. Yu, L. Diosi, N. Gisin, and W. T. Strunz, Phys. Rev. A
\textbf{60}, 91 (1999).

\bibitem {YuQBM}W. T. Strunz and T. Yu, Phys. Rev. A \textbf{69}, 052115 (2004).

\bibitem {Jing-Yu2010}J. Jing and T. Yu, Phys. Rev. Lett. \textbf{105}, 240403 (2010).



\bibitem {JunArxiv}J. Jing, \textit{et al}., arXiv:1012.0364 (to be published).



\bibitem {Xinyu2011}X. Zhao \textit{et al}., Phys. Rev. A \textbf{84}, 032101 (2011).

\bibitem {Broadbent2011}C. J. Broadbent \textit{et al}., Arxiv preprint
arXiv:1112.2716 (2011).

\bibitem {Chen}H. Yang, H. Miao, and Y. Chen, arXiv:1108.0963v3 (2011).

\bibitem {JingEPL}J. Jun and T. Yu, Eruo. Phys. Lett. \textbf{96} 44001 (2011).

\bibitem {Eisfeld2011}G. Ritschel \textit{et al}., arXiv:1108.3452v2.

\bibitem {1qu-in-BF}L.-D. Chang and S. Chakravarty, Phys. Rev. B \textbf{31},
154 (1985).

\bibitem {Search2002}C. P. Search \textit{et al}., Phys. Rev. A \textbf{66},
043616 (2002).

\bibitem {Sinha2010}S. S. Sinha \textit{et al}., Phys. Rev. E \textbf{82},
051125 (2002).

\bibitem {QBM-Fbath}E. S. Hern\'{a}dez and C. O. Dorso, Phys. Rev. C
\textbf{29}, 1510 (1984).

\bibitem {1quFgas}K. Vlad\'{a}rand G. T. Zim\'{a}nyi Phys. Rev. Lett.
\textbf{56}, 286 (1986).

\bibitem {Wufu2011}W. Shi, X. Zhao, and T. Yu, (to be published).

\bibitem {Alvermann2008}A. Alvermann and H. Fehske, Phys. Rev. B. \textbf{77},
045125 (2008).

\bibitem {Barouch1970}E. Barouch, B. McCoy, and M. Dresden, Phys. Rev. A
\textbf{2,} 1075 (1970).

\bibitem {ZhangRMP}W. -M. Zhang, D. H. Feng, and R. Gilmore, Rev. Mod. Phys.
\textbf{62}, 867 (1990).

\bibitem {Glauber1999}K. E. Cahill and R. J. Glauber, Phys. Rev. A
\textbf{59}, 1538 (1999).

\bibitem {Yu-FiniteT}T. Yu, Phys. Rev. A \textbf{69}, 062107 (2004).

\bibitem {comment}Here we still use the terminology environment even it
contains only a few degrees of freedom.
\end{thebibliography}
\end{document}